\documentstyle[12pt]{article}

\begin{document}
\title{Multifractal Analysis of the Coupling Space of Feed-Forward
Neural Networks}
\author{ A.~Engel and M. Weigt \\[.3cm]
      Institut f\"ur Theoretische Physik \\
      Otto-von-Guericke-Universit\"at Magdeburg\\
      PSF 4120\\
      39016 Magdeburg, Germany}

\date{July, 13, 1995}
\maketitle
\begin{abstract}
Random input patterns induce a partition of the coupling space of
feed-forward neural networks into different cells according to the
generated output sequence. For the perceptron this partition forms a random
multifractal for which the spectrum $f(\alpha)$ can be calculated
analytically using the replica trick. Phase transition in the
multifractal spectrum correspond to the crossover from percolating to
non-percolating
cell sizes. Instabilities of negative moments are related to the
VC-dimension.

\vspace{.5cm}\noindent
\end {abstract}

Multifractal concepts were originally introduced in the context of
developed turbulence \cite{Man} and
chaotic dynamical systems \cite{Grass} and have
become since then a standard tool to analyze physical systems with
richer structure than that induced by dilation symmetry alone (for
reviews see \cite{PV}).
Contrary to simple scale invariant situations as provided, e.g.,
by systems at second order phase transitions
that can be classified with the help of a few critical exponents only the
description of multifractals reqires a full range of scaling
exponents specified by a continuous function $f(\alpha)$. The reason
for this multitude of exponents is the fact that different
moments of the underlying probability distribution are dominated by
different fractal subsets of the system.

The simplest examples of multifractal measures are provided by
deterministic recursive constructions as the two-scale Cantor set
\cite{PV}. However, many multifractals observed experimentally
as, e.g., in diffusion limited aggregation, are generated by
random processes. It is therefore of general interest to analyze simple
models of random systems that exhibit multifractality \cite{JPV}. A
particularily simple case is provided by fractals on which a measure
of constant density is distributed. The multifractal properties are
then of purely geometrical origin and characterize the fractal
support itself \cite{TeVi}.

In the present letter we show that the coupling space of simple
feed-forward neural networks storing random input-output mappings
displays multifractality. The corresponding spectrum $f(\alpha)$ can
be determined explicitely using the replica trick. On the one hand
these systems may hence serve as examples to test the properties of random
multifractals. On the other hand the multifractal analysis of the
cell structure imposed on the coupling space by the random
input-output mappings refines and extends the standard statistical
mechanics analysis of the storage \cite{Ga} and
generalization properties \cite{GT} of these systems.

We consider a perceptron with $N$ input bits $\xi_i =\pm 1, i=1,...,N$, and
one output $\sigma=\pm 1$. The output is given as the sign of the
scalar product between the input and the coupling vector {\boldmath $J$}
of the perceptron, i.e.
\begin{equation}\label{1}
\sigma=\mbox{sign}(\mbox{\boldmath $J$} \mbox{\boldmath $\xi$})
      =\mbox{sign}(\sum_i J_i \xi_i)
\end{equation}
We will mainly consider the case
in which the \mbox{\boldmath $J$}-vector is binary, $J_i=\pm 1$,
(Ising-perceptron) and
defer the discussion of continuous couplings (spherical perceptron) to
the end of this letter.

Given $p=\gamma N$ different input patterns $\mbox{\boldmath $\xi$}^{\mu}$
we can associate
with every pattern a hyperplane perpendicular
to it that cuts the coupling space of the perceptron into two
halfs according to the possible output $\sigma^{\mu}=\pm 1$. If the
inputs are generated independently at random we will hence find a
{\sl random partition of the coupling space} into at most $2^p$ cells. These
cells can be labeled by their output sequences
$\mbox{\boldmath $\sigma$}=\{\sigma^{\mu}\}$ and their size gives the
probability
$P(\mbox{\boldmath $\sigma$})$ that, for given input sequence
$\{\mbox{\boldmath $\xi$}^{\mu}\}$,
the outputs $\mbox{\boldmath $\sigma$}$ are generated by a randomly chosen
coupling vector \mbox{\boldmath $J$}. It is well known that the storage and
generalization properties of the neural network are closely related
to this probability distribution \cite{Cov,HKS,DGP}.

The natural scale for the cell sizes in the thermodynamic limit
$N\to\infty$ is $\epsilon =2^{-N}$.
Due to the random orientation of the hyperplanes, however, the cells
will differ significantly in size from each other.
To describe these fluctuations quantitatively we introduce the crowding
index $\alpha(\mbox{\boldmath $\sigma$})$ by
\begin{equation}\label{2}
   P(\mbox{\boldmath $\sigma$})=\epsilon^{\alpha(\mbox{\boldmath $\sigma$})}
\end{equation}
and characterize $P(\mbox{\boldmath $\sigma$})$ by its moments
\begin{equation}\label{3}
   \langle P^q \rangle =\sum_{\mbox{\boldmath $\sigma$}} P^q(\mbox{\boldmath
$\sigma$})=\epsilon^{\tau(q)}
\end{equation}
with the mass exponent $\tau(q)$. As usual \cite {Man,Grass} the
multifractal spectrum $f(\alpha)$ is given by the Legendre transform
of $\tau(q)$:
\begin{equation}\label{4}
   f(\alpha)=\min_q[\alpha q - \tau(q)]
\end{equation}
and ${\cal N}(\alpha) = \epsilon^{f(\alpha)}$ gives the number of cells of
size $\epsilon^{\alpha}$.

For large $N$ we expect that $\tau$ and $f$ become self-averaging,
i.e. they will no longer depend on the choice of
the random inputs $\mbox{\boldmath $\xi$}^{\mu}$.
We can therefore calculate $\tau(q)$ from
\begin{equation}\label{5}
\tau(q)=-\lim_{N\to\infty}\frac{1}{N \ln
2}\langle\langle\ln\sum_{\mbox{\boldmath $\sigma$}}
         P^q(\mbox{\boldmath $\sigma$})\rangle\rangle
\end{equation}
where $\langle\langle ...\rangle\rangle$ denotes the average over the
distribution of the input patterns which we take as
\begin{equation}\label{6}
{\cal P}(\xi_i^{\mu})=\prod_{i,\mu}[\frac{1}{2}\delta(\xi_i^{\mu}+1)+
                               \frac{1}{2}\delta(\xi_i^{\mu}-1)]
\end{equation}
The calculation of $\tau(q)$ uses a variant of the replica trick introduced
recently by Monasson and O'Kane \cite{MoKa} and will only be
sketched. Starting with the definition of $P(\mbox{\boldmath $\sigma$})$
\begin{equation}\label{7}
 P(\mbox{\boldmath $\sigma$})=\sum_{\mbox{\boldmath $J$}}
\prod_{\mu=1}^p \theta(\sigma^{\mu} \mbox{\boldmath $J$}
\mbox{\boldmath $\xi$}^{\mu})
\end{equation}
with the theta-function $\theta(x)=1$ if $x\geq 0$ and $\theta(x)=0$
else we introduce one replica index $\alpha$ running from $1$ to $q$ in
order to represent the $q$-th power of $P$ and another replica index
$a$ running from $1$ to $n$ to represent the ln in eq.(\ref{5}) in
the usual way \cite{MPV}. We then find
\begin{equation}\label{8}
\tau(q)=-\lim_{N\to\infty}\frac{1}{N \ln 2}\lim_{n\to 0}\frac{1}{n}
         \left[ \langle\langle\sum_{\sigma_{\mu}^a}\sum_
         {\mbox{\boldmath $J$}^{a \alpha}}
         \prod_{\mu,a,\alpha} \theta(\sigma_{\mu}^a
         \mbox{\boldmath $J$}^{a\alpha} \mbox{\boldmath $\xi$}^{\mu})
         \rangle\rangle -1 \right]
\end{equation}
Next the average is performed and the resulting expression is written
as a saddle-point integral over the elements of the overlap matrix
\begin{equation}\label{9}
  Q_{ab}^{\alpha\beta}=\frac{1}{N}\sum_i J_i^{a \alpha} J_i^{b \beta}
\end{equation}
and its conjugate $\hat{Q}_{ab}^{\alpha\beta}$. To
solve the remaining extremalization problem we assume replica symmetry
(RS) which in the present problem is given by the ansatz
\cite{MoKa}
\begin{eqnarray}\label{10}
  Q_{ab}^{\alpha\beta}=Q_1 & \mbox{if} &
         a=b,\alpha\neq\beta\\\nonumber
  Q_{ab}^{\alpha\beta}=Q_0 & \mbox{if}& a\neq b
\end{eqnarray}
and similarily for $\hat{Q}_{ab}^{\alpha\beta}$. Hence $Q_1$ denotes
the typical overlap between couplings in the same cell (same output
vector \mbox{\boldmath $\sigma$}) whereas $Q_0$ characterizes the overlap
between couplings
in different cells. The reliability of the RS-ansatz will be
discussed below.

In this way we get
\begin{eqnarray}\label{11}
   \tau(q)&=&\frac{1}{\ln 2}\, \mbox{extr}_{Q_1,Q_0,\hat{Q}_1,\hat{Q}_0}
       \left[\frac{q\hat{Q}_1}{2}(1-(1-q)Q_1)-\frac{q^2}{2}\hat{Q}_0 Q_0
       \right.\\\nonumber
   & & \left. - \int Dz_0\ln\int Dz_1 \cosh^q
       (\sqrt{\hat{Q}_1-\hat{Q}_0}z_1 + \sqrt{\hat{Q}_0}z_0)\right.\\\nonumber
   & & \left. - \gamma\int Dt_0\ln 2\int Dt_1
       H^q(\frac{\sqrt{Q_1-Q_0}t_1+\sqrt{Q_0}t_0}{\sqrt{1-Q_1}})\right]
\end{eqnarray}
where $Dx=\exp(-x^2/2)/\sqrt{2\pi}$ and $H(y)=\int_y^{\infty} Dx$.
Inspection of the saddle-point equations
resulting from (\ref{11}) reveals that these are always fulfilled for
$Q_0=\hat{Q}_0=0$. Physically this is due to the symmetry
$(\mbox{\boldmath $J$},\mbox{\boldmath $\sigma$})
\leftrightarrow (-\mbox{\boldmath $J$},-\mbox{\boldmath $\sigma$})$
in eq.(\ref{1}) which ensures
that to every cell there is a "mirror"-cell of equal size. As a consequence
eq.(\ref{11}) simplifies to
\begin{eqnarray}\label{12}
   \tau(q)&= &\frac{1}{\ln 2}\,\mbox{extr}_{Q_1,\hat{Q}_1}\left[
       \frac{q\hat{Q}_1}{2}(1-(1-q)Q_1)\right.\\\nonumber
   & & \left.-\ln\int Dz \cosh^q(\sqrt{\hat{Q}_1}z)-\gamma\ln 2\int Dt
       H^q(\frac{\sqrt{Q_1}t}{\sqrt{1-Q_1}})\right]
\end{eqnarray}
Extremizing this expression with respect to $Q_1$ and $\hat{Q}_1$
numerically and using eq.(\ref{4}) we find $f(\alpha)$ as
shown in the figure for different values of the loading parameter $\gamma$.
Also shown are results from exact enumerations up to
$N=30$. The inset gives a finite size analysis demonstrating very
good agreement between the analytical results and the extrapolation
from the numerical data.

For small values of $\gamma$ the $f(\alpha)$-curves
have the typical bell-shaped form. The two zeros $\alpha_{min}(\gamma)$ and
$\alpha_{max}(\gamma)$
specify the largest \cite{BiOp} and the smallest cell occuring with
non-zero probability respectively.
Moreover $\alpha_0(\gamma)= $argmax$f(\alpha)$
defines the size of the typical cell and $f(\alpha_0)=\gamma$ indicates that
for small $\gamma$ all possible cells do indeed occur.
Finally, from the normalization of $P(\mbox{\boldmath $\sigma$})$ we find for
all $\gamma$
that $\tau(q=1)=0$. This implies that the $f(\alpha)$-curves are all
tangent to the line $f=\alpha$. We denote the abcissa of the tangential
point by $\alpha_1(\gamma)$. Cells of size $\epsilon^{\alpha_1}$ contribute
most to the coupling space.

For larger values of $\gamma$ it becomes important that due to
$J_i=\pm 1$ the coupling space is discrete and the cells sizes
must always be multiples of $\epsilon$. Values of $\alpha$ larger than $1$
thus correspond to empty cells. For $\alpha_0=1$ the {\sl typical} cell is
empty and hence $\alpha_0(\gamma_c)=1$ determines the storage capacity
$\gamma_c$ \cite{DGP}. From the figure we infer $\gamma_c\cong .833$.
Similarily $\alpha_1 =1$ indicates
that the coupling space is dominated from cells containing a single
\mbox{\boldmath $J$}-vector only.
Therefore $\alpha_1(\gamma_g)=1$ defines the threshold to
perfect generalization \cite{DGP} and the results shown in the figure yield
$\gamma_g=1.245$.

Both the values for $\gamma_c$ and $\gamma_g$ have
been derived previously \cite{KM} and in fact $\alpha(\gamma)=0$ is similar
to the zero-entropy condition used to derive them. However, it should be
emphasized that the justification of the zero-entropy condition within the
traditional
Gard\-ner-approach requires one-step replica symmetry breaking (RSB).
Contrary, in the present approach not only RS already gives the
correct results but anticipating that $Q_0=\hat{Q}_0=0$ for
$\alpha_1<\alpha<\alpha_0$ (see below) we can even go without the
replica index $a$ and calculate $\tau(q)$ as an {\sl annealed}
average over the input distribution (\ref{6}). This is technically
much simpler than a one-step RSB calculation and may open the way to
a more detailed analysis of multilayer networks also \cite{MoZe}.

We now turn to the reliability of the RS results with $Q_0=0$.
First of all we note that from (\ref{4}) and (\ref{3}) $f(\alpha)$ is
the entropy of the discrete spin system \mbox{\boldmath $\sigma$}
(with hamiltonian $\alpha(\mbox{\boldmath $\sigma$})$)
and must therefore be
non-negative. However, before $f$ becomes negative we find for both
positive and negative $q$ a transition to a saddle point with
$Q_0,\hat{Q}_0 > 0$  at values of $q=q_{\pm}$ given by
\begin{equation}\label{13}
   \sqrt{\alpha}=|q_{\pm}-1|\hat{Q}_1(1-Q_1)(1-(1-q_{\pm})Q_1)
\end{equation}
Since $Q_0$ gives the typical overlap between \mbox{\boldmath $J$}-vectors
belonging to
different cells the analogy with the spin glass problem suggests
that $Q_0>0$ signals broken ergodicity. This means in the
present context that moments of $P(\mbox{\boldmath $\sigma$})$ with $q>q_+$ or
$q<q_-$ are dominated
by cells that no longer percolate in coupling space. Whereas for
$q_+>q>q_-$ all dominating cells can be reached from one another without
entering cells of a different size this is no longer true for the
remaining values of $q$.
Note that the transition occurs always outside the interval
$(\alpha_1,\alpha_0)$.

We have also investigated the transversal stability of the RS saddle
point using standard techniques \cite{AT}. We find that at
$q=q_{\pm}$ the RS saddle point becomes unstable also with respect to RSB
and this instability is not removed by $Q_0>0$ \cite{rem}.
On the other hand we believe that our qualitative
picture of a percolation transition at $q_c$ remains valid also in
a RSB solution.

We have obtained similar results for the spherical perceptron (see also
\cite{BiOp}). For small
values of $\gamma$ the $f(\alpha)$-curves are almost identical to
those of the Ising-perceptron. Since there is now no
smallest possible size of a cell the storage capacity $\gamma_c=2$
\cite{Cov,Ga} has to be determined from $\alpha_0(\gamma_c)\to\infty$.
For $\gamma\geq\gamma_c$ the spectrum $f(\alpha)$ is hence monotonously
increasing and the asymptotic value of $f$ for $\alpha\to\infty$ remains
smaller than $\gamma$ since a growing fraction of classifications
cannot be realized. The generalization properties of the spherical
perceptron are characterized by the information dimension $f(\alpha_1)$
of the multifractal cell structure which is related to the volume
${\cal V}$ of the version space by ${\cal V}=\epsilon^{f(\alpha_1)}$
\cite{DGP}. The longitudinal and transversal instability of the
RS solution with $Q_0=0$ occurs again at the same values $q_{\pm}$ of
$q$ given now by $\sqrt{\alpha}=|q_{\pm}-1| Q_1$.

Finally we find from the RS results for both the Ising and the
spherical perceptron an
instability $\langle P^q \rangle \to\infty$ of negative moments if
$q<q_c$ where $q_c=1-1/\alpha$ for $\alpha\leq 1$ and $q_c=0$ for
$\alpha>1$. The corresponding endpoint of the $f(\alpha)$-curve
for $\gamma=.4$ is marked by a dot in the figure, for $\gamma=.2$ it
occurs for $f<0$.
These divergencies are due to the
possibility of empty cells with $P(\mbox{\boldmath $\sigma$})=0$.
For small values of $\gamma$ we find $q_c<q_-$ and the instability
lies outside the region of validity of RS. In this case it is
therefore necessary to include RSB to elucidate the nature of this
divergence and we speculate that this is due to the fact that empty
cells can only occur due to very rare realizations of the input patterns
$\mbox{\boldmath $\xi$}^{\mu}$ \cite{rem2}.
For $\gamma$ larger than a threshold value $\gamma^{VC}$, however, the
instability occurs within the region of validity of RS. Then the
probability of empty cells can no longer be exponentially small in $N$.
The instability of negative moments is hence related to
the Vapnik-Chervonenkis dimension of the neural network, more
precisely $\gamma^{VC}$ determined from $q_c(\gamma^{VC})=q_-(\gamma^{VC})$
gives
an upper bound on the VC-dimension $d_{VC}$ \cite{VC}.
For the spherical perceptron we find in this way the well known exact result
$\gamma^{VC}=1$ \cite{Cov}. For the Ising perceptron we get similarily
$\gamma^{VC}\cong 0.557$. The VC-dimension of the Ising
perceptron is not known up to now. One can show that it is at
least 0.5 \cite{Helm} and there is numerical evidence that it is
indeed 0.5 for large $N$ \cite{Sta}. Our RS upper bound is
somewhat larger which may indicate that there is a {\sl discontinuous}
transition to RSB for the Ising perceptron. This question is
currently under study.

In conclusion we have shown that a multifractal analysis of the
phase space of neural networks allows to refine the standard
statistical mechanics approach to learning and generalization substantially.
The instabilities found in the spectrum $f(\alpha)$ can
be related to physical properties of these systems. We hope that these
results may serve as a guide
to understand similar transitions in other examples of random multifractals.
Moreover it would be interesting to see whether there are other examples for
percolation transitions in high-dimensional spaces that can be detected from
a bifurcation of an overlap parameter $Q_0$.

\section*{Acknowlegdement}
We would like to thank Chris van den Broeck for bringing the connection between
the
cell structure of the perceptron and multifractals to our attention.
Stimulating discussions with  G\"unther Radons and
Manfred Opper are gratefully acknowledged.

\newpage
\section*{Figure caption}
\bigskip
\begin{quote}
Multifractal spectrum $f(\alpha)$ characterizing the cell structure of
the coupling space of an Ising-perceptron classifying $\gamma N$
random input patterns for $\gamma=0.2, 0.4, 0.833$, and $1.245$ (from
left to right). The diamonds mark the region of validity of the replica
symmetric ansatz. The dot denotes the location of the instability of
negative moments for $\gamma=.4$ (see text).
The histograms are exact enumeration results for
$N=30, p=6; N=30, p=12$ and $N=24, p=20$ respectively. The inset
shows a finite size analysis of $f(\alpha_0)$ for $\gamma=0.5$. The
correct value for $N\to\infty$ is $0.5$, the line describes the first
correction to the saddle point for $N<\infty$, and the statistical
error of the numerical results is smaller than the symbol size.
\end{quote}


\begin{thebibliography}{[99]}
\bibitem{Man} B. B. Mandelbrot, J. Fluid. Mech. {\bf 62}, 331 (1974),
    U. Frisch and G. Parisi, in {\sl Turbulence and
    Predictability in Geophysical Fluid Dynamics and Climate Dynamics},
    M. Ghil, R. Benzi, G. Parisi (eds.) (North Holland, 1985),
    R. Benzi, G. Paladin, G. Parisi, and A. Vulpiani
    J. Phys. {\bf A17}, 3521 (1984)
\bibitem{Grass} P. Grassberger, Phys. Lett. {\bf 97A}, 227 (1983),
    T. C. Halsey, M. H. Jensen, L. P. Kadanoff, I. Procaccia,
    and B. I. Shraiman, Phys. Rev. {\bf A33}, 1141 (1986)
\bibitem{PV} G. Paladin, A. Vulpiani, Phys. Reports {\bf 156}, 147
    (1987), J. Feder, {\sl Fractals} (Plenum, New York, 1988),
    A. Bunde and S. Havlin (eds.) {\it Fractals and
    Disordered Systems}, (Springer, Berlin, 1991)
\bibitem{JPV} M. H. Jensen, G. Paladin, A. Vulpiani, Phys. Rev.
    {\bf E50}, 4352 (1994), Y. Y. Goldschmidt and T. Blum, Phys. Rev. {\bf
     E48}, 161 (1993)
\bibitem{TeVi} T. T\'{e}l and T. Vicsek, J. Phys. {\bf A20}, L835 (1987)
\bibitem{Ga} E. Gardner, J. Phys. {\bf A21}, 257 (1988),
    E. Gardner and B. Derrida, J. Phys. {\bf A21}, 271 (1988)
\bibitem{GT} G. Gy\"orgyi and N. Tishby, {\it Workshop on Neural Networks
    and Spin Glasses} K.~Theumann and W.~K.~Koeberle (eds.), (World
    Scientific, Singapore, 1990), M. S. Seung, H. Sompolinsky and N. Tishby,
    Phys. Rev. {\bf A45}, 6056 (1992)
\bibitem{HKP} J. A. Hertz, A. Krogh, and R. G. Palmer,
    {\it Introduction to the theory of neural
    computation},(Addison-Wesley, Redwood City, 1991)
\bibitem{Cov} T. M. Cover, IEEE Trans. Electron. Comput. {\bf EC-14},
    326, (1965)
\bibitem{HKS} D. Haussler, M. Kearns, and R. Schapire, {\it Bounds on
    the Sample Complexity of Bayesian Learning Using Information Theory
    and the VC Dimension}, Proceedings {\sl COLT '91}, Morgan Kaufmann,
    San Mateo., M. Opper, D. Haussler, Phys. Rev. Lett. {\bf 66}, 2677,
    (1991)
\bibitem{DGP} B. Derrida, R. B. Griffith, and A. Pr\"ugel-Bennett,
    J. Phys. {\bf A24}, 4907 (1991)
\bibitem{MoKa} R. Monasson and D. O'Kane, Europhys. Lett. {\bf 27},
    85 (1994)
\bibitem{MPV} M.~Mezard, G.~Parisi, M.~A.~Virasoro  {\it Spin glass
    theory and beyond} (World Scientific, Singapore, 1987)
\bibitem{BiOp} M. Biehl and M. Opper, {\it Perceptron Learning: The
    largest Version Space}, Preprint, Universtity of W\"urzburg, 1995
\bibitem{KM} W. Krauth and M. Mezard, J. Phys. France {\bf 50}, 3057
    (1989), G. Gy\"orgyi, Phys. Rev. Lett. {\bf 64}, 2957 (1990)
\bibitem{MoZe} R. Monasson and  R. Zecchina,{\it Weight space
    structure and internal representations: A Direct approach to Learning
    and Generalization in multilayer neural nets}, Preprint, Rome
    University, 1995
\bibitem{AT} J. R. L. de Almeida, D. Thouless, J.Phys. {\bf A11}
    983 (1978), T. Temesvari, C. De Dominicis, and I. Kondor,
    J. Phys. {\bf A27}, 7569 (1994)
\bibitem{rem} This is reminiscent of the SK-model in zero field. The
    analogue of a non-zero field situation can be studied by introducing
    a threshold of the perceptron.
\bibitem{rem2} For large N the probability for patterns
    distributed according to (\ref{6})
    not to be in general position scales as $2^{-N}$.
\bibitem{VC} V. N. Vapnik and A.Y. Chervonenkis, Th. Prob. Appl.
    {\bf16}, 264 (1971), A. Engel, Modern Physics Letters {\bf B8}, 1683 (1994)
\bibitem{Helm} D. Helmbold, private communication,
\bibitem{Sta} G. Stambke, Diploma Thesis, University of Giessen, 1992
\end{thebibliography}
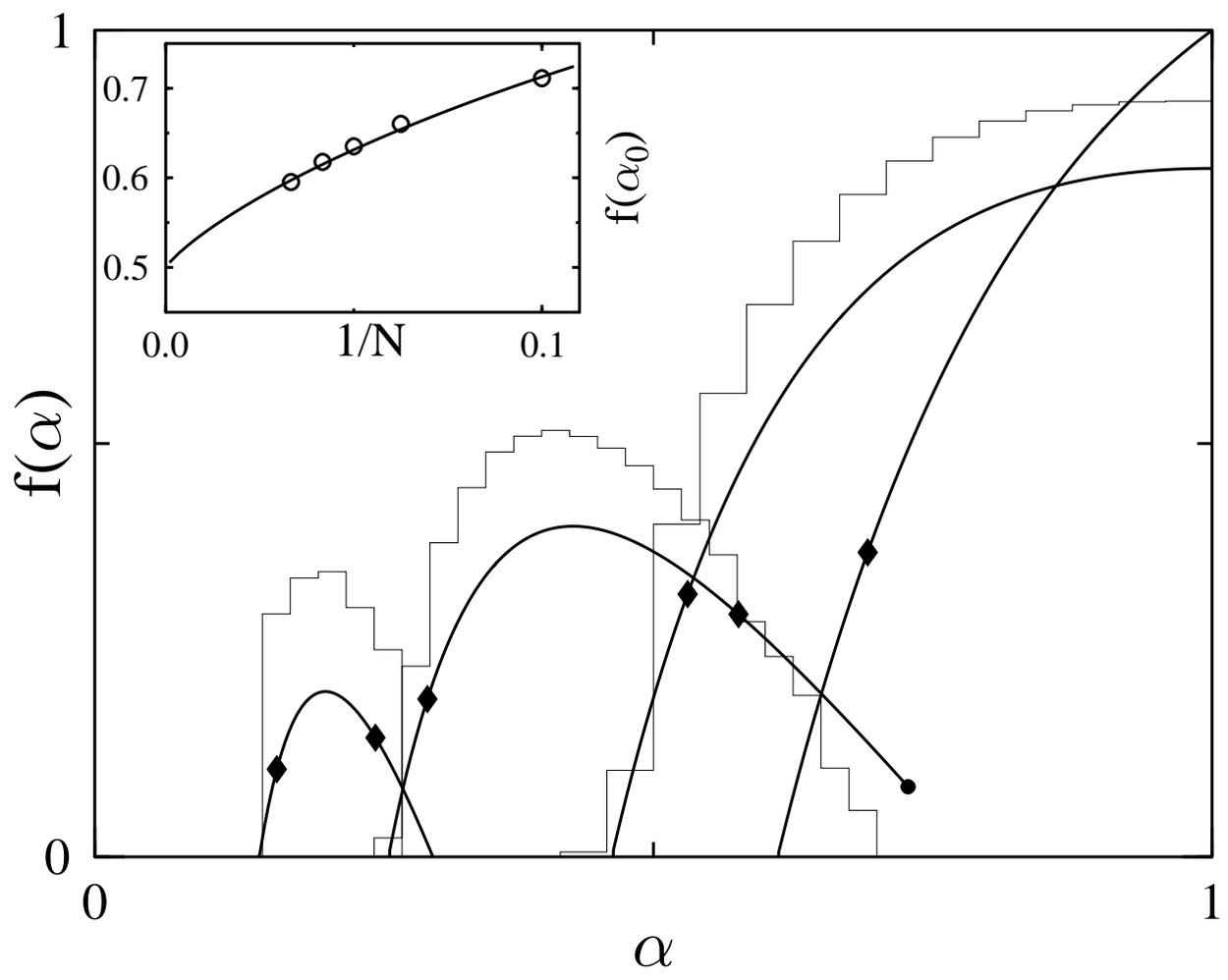
\end{document}